\documentclass[twocolumn,showpacs,preprintnumbers,amsmath,amssymb]{revtex4}

\usepackage{CJK}

\usepackage{graphicx}

\usepackage{dcolumn}

\usepackage{bm}
\usepackage[normalem]{ulem}
\usepackage{color}

\usepackage{epstopdf}

\def\rf#1{(\ref{#1})}

\newcommand{\eps}{\epsilon}

\newcommand{\bew}{\begin{widetext}}
\newcommand{\ew}{\end{widetext}}
\newcommand{\nn}{\nonumber}

\newcommand{\ed}{\eta_{_\Delta}}
\newcommand{\etg}{\eta_{_\Gamma}}

\newcommand{\br}{\mathbf{r}}
\newcommand{\Nb}{\langle N \rangle}

\newcommand{\bq}{\mathbf{q}}
\newcommand{\bv}{\mathbf{v}}

\newcommand{\bR}{\mathbf{R}}

\newcommand{\sep}{ \ \ \ , \ \ \ }

\newcommand{\beq}{\begin{equation}}
\newcommand{\eeq}{\end{equation}}
\newcommand{\beqn}{\begin{eqnarray}}
\newcommand{\eeqn}{\end{eqnarray}}

\newcommand{\dd}{{\rm d}}
\newcommand{\ee}{{\rm e}}

\usepackage{bm}
\usepackage[normalem]{ulem}

\def\beq{\begin{equation}}
\def\eeq{\end{equation}}
\def\bea{\begin{eqnarray}}
\def\eea{\end{eqnarray}}
\def\rf#1{(\ref{#1})}

\begin{document}

\begin{CJK*}{GBK}{}

\title{ Only the immortal can (almost)  flock while turning: Two-dimensional chiral immortal flocks, time  crystals and  non-universal sub-logarithmic roughness}
\author{Sudip Mukherjee}\email{sudip.bat@gmail.com}
\affiliation{Barasat Government College,
10, KNC Road, Gupta Colony, Barasat, Kolkata 700124,
West Bengal, India}
\affiliation{Max-Planck Institut f\"ur Physik Komplexer Systeme, N\"othnitzer
Str. 38,
01187 Dresden, Germany}
\author{Abhik Basu}\email{abhik.basu@saha.ac.in,abhik.123@gmail.com}
\affiliation{Theoretical Physics Division, Saha Institute of
Nuclear Physics, Calcutta 700064, West Bengal, India}
\affiliation{Max-Planck Institut f\"ur Physik Komplexer Systeme, N\"othnitzer
Str. 38,
01187 Dresden, Germany}
\author{Leiming Chen}
\email{leiming@cumt.edu.cn}
\affiliation{School of Material Science and Physics, China University of Mining and Technology, Xuzhou Jiangsu, 221116, P. R. China}
\author{John Toner}\email{jjt@uoregon.edu}
\affiliation{Department of Physics and Institute of Theoretical Science, University of Oregon, Eugene, Oregon 97403, USA}
\affiliation{Max-Planck Institut f\"ur Physik Komplexer Systeme, N\"othnitzer
Str. 38,
01187 Dresden, Germany}

%\end{CJK*}

\date{\today}
\begin{abstract}
We study two-dimensional chiral ``dry''  (non-momentum conserving) ``immortal'' { (number conserving)} ``flocks'' { (polar ordered active fluids)}.
Noise driven fluctuations about the noise-free
{ ``time crystal''}  state, in which
the velocity rotates uniformly in time,
obey coupled  Kardar-Parisi-Zhang-like equations  for
 { the local velocity orientation and a conserved}
concentration field; the latter field
 stabilizes  order. Orientational  fluctuations grow only sub-logarithmically, and
number fluctuations are suppressed (``hyperuniform'');  both exhibit
 nonuniversal scaling exponents.
 There are three different universality classes of disordering transitions.
 \end{abstract}

%\pacs{05.65.+b, 64.60.Ht, 87.18Gh}
\maketitle
\end{CJK*}

 Flocking - the collective, coherent motion of a large number of active particles - is { the canonical } example of active matter~\cite{JT-book,SR-AnnRev,sriram-RMP,vicsek,toner-tu-prl,toner-tu-pre,JT-annphys}.  { The }  most common { examples  occur in biological systems, where they are found}  over an enormous range of length scales: from herds of caribou, through flocks of birds and schools of fish, down to bacteria colonies and even the biomolecular level, e.g., the mitotic spindle~\cite{spindle,spindle1}. Synthetic %non-biological
 examples (e.g., ``flocks'' of quincke rotators~\cite{bartolo}) have also been studied.

 Flocking is particularly striking in two dimensions (2D), since, in order to move coherently, a flock must spontaneously break continuous rotation invariance, an impossibility in 2D {\it equilibrium} systems~\cite{mwt,mwt1}. Thus, the order in  { 2D} flocks is an intrinsically non-equilibrium phenomenon
  ~\cite{JT-book,vicsek,toner-tu-prl,toner-tu-pre,JT-annphys}.

 Besides activity, {\em chirality}, i.e.,  the absence of left-right symmetry, is also widespread in biology.  Most earlier studies\cite{JT-book,SR-AnnRev,sriram-RMP,vicsek,toner-tu-prl,toner-tu-pre,JT-annphys} have focused only on achiral flocks, which retain left-right symmetry. %This raises the question of whether chirality can destroy the robust long-range orientational ordered 2D achiral flocks survives.
 Among the few exceptions are numerical studies of
 simple chiral modifications of the Vicsek model,
 %\sout{can simulate chiral flocks and produce}
  which find chiral flocking states~\cite{Chate_chiral},  and  recent   analytic studies~\cite{JT-chiral-pre}  of non-number-conserving (``Malthusian'') flocks, in which active agents are continuously born and die. The latter study  showed, by mapping these systems to the 2D KPZ equation~\cite{kpz},   that chirality destroys long-range orientational order  in chiral Malthusian flocks,  leaving
 %\sout{By mapping these systems to the 2D KPZ equation~\cite{kpz}, researchers found that chirality leads to}}
 only short-range order,  rather than  either the quasi-long-range order seen in 2D equilibrium XY models~\cite{chaikin},  or the long-range order found in achiral 2D flocks.

 In this Letter, we study immortal chiral dry flocks, with a fixed number of active agents moving on a frictional substrate, i.e., momentum is not conserved. Number conservation is the system's only conservation law.

We find that dry chiral immortal  flocks, while still not truly long-ranged orientationally ordered, have a phase in which they are {\it better} ordered than 2D equilibrium XY models. This phase is also ``hyperuniform''; i.e., it has
miniscule number fluctuations.

This ``better ordered'' chiral flock can be understood by comparing it to an ideal, perfectly ordered one. Because chirality causes all agents to spontaneously turn in the same direction (left or right) as they move, the ideal state~\cite{Chate_chiral,chiral-refs1,chiral-refs2} has a coarse-grained velocity field ${\bf v}({\bf r},t)$ that is uniform in space  but periodic in time~\cite{JT-chiral-pre}:
\beq
\bv=v_0[\cos(bt+\phi)\hat{\bf x}-\sin(bt+\phi)\hat{\bf y}] \,.%,,~~~ \rho=\rho_0,
\label{SSI}
\eeq
Here the parameter $b$ is an intrinsic frequency characteristic of the system, whose sign  determines whether the agents spontaneously turn to the left ($b>0$) or the right ($b<0$).    An achiral system must have $b = 0$. Hence,  $b$ may be used as a measure of the chirality
of the system.
Here, $v_0$  is the mean speed of the flockers, and $\phi$ a  phase which is spontaneously chosen by the system. Since $\bv$ in (\ref{SSI})  is spatially uniform at any $t$,  the state \rf{SSI} has infinite-ranged orientational order.  This  state  can be thought of as a ``time crystal'';  or, more precisely, as a
%\sout{, namely
``time cholesteric", since
the velocity rotates in time in the same way that the molecular axis rotates in
space in a three-dimensional cholesteric liquid crystal~\cite{deGennes}.

\begin{figure}[htb]
\includegraphics[width=\columnwidth]{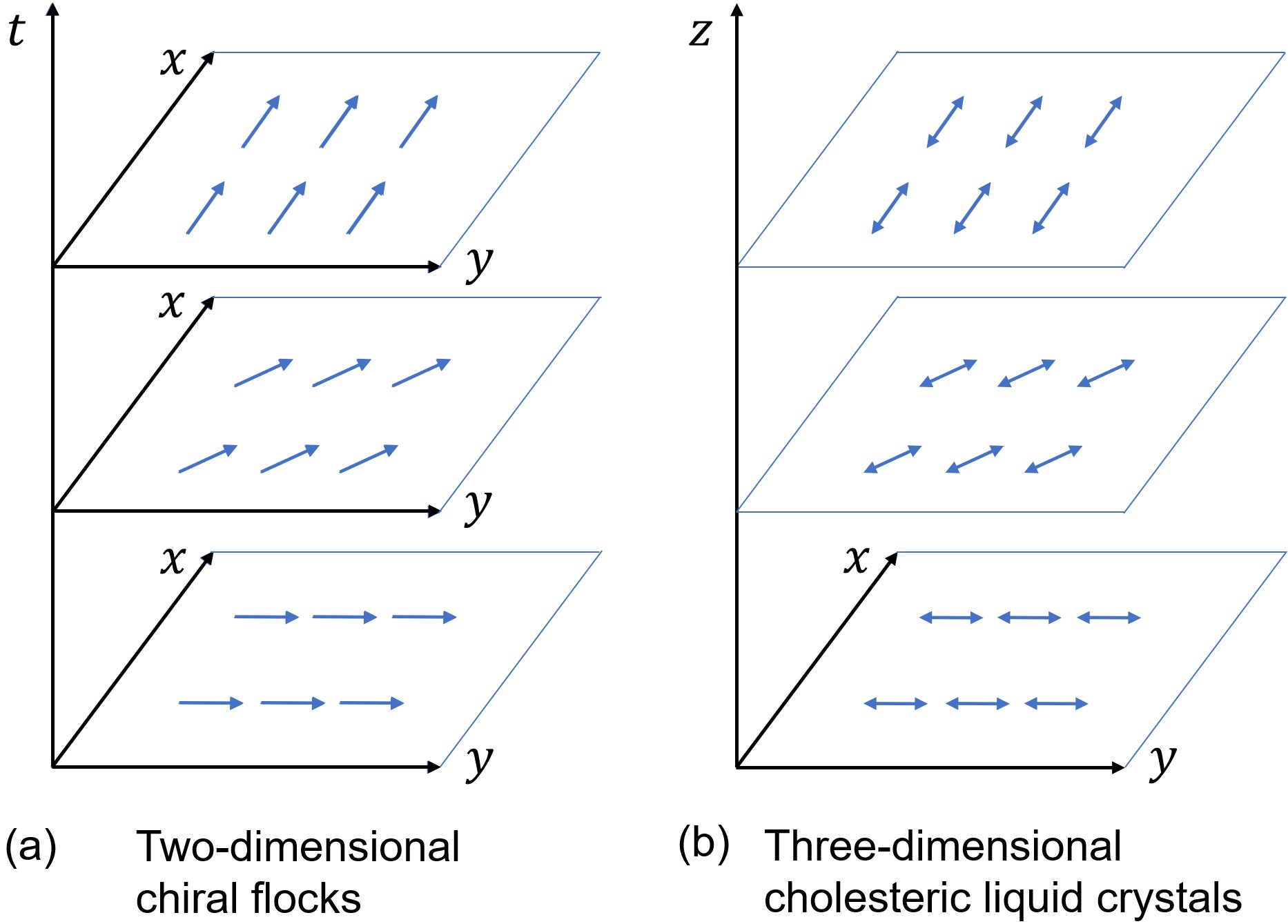}%\hfill \includegraphics[width=0.4\columnwidth]{Phase region_final.jpg}
 \caption{The analogy between 2D chiral flocks (left) and three-dimensional cholesteric liquid crystals (right). The single-head arrows represent polar active particles with their velocities pointing in the directions of the arrows. The double-headed arrows represent apolar rod-like long molecules of cholesteric liquid crystals. %(b)Phase diagram of the model in the $\mu_1-\mu_2$ plane{\cmag .\sout{ in the linearly stable sector.} %\sout{The blue region represents the}
 %{\cblue Blue: sublogarithmic-rough regime with $0<\eta<1$; green: %\sout{the green one the super-rough regime}
 %speculated short-ranged order due to vortex unbinding,\sout{ with $\eta>1$}, red: unstable region with negative diffusivities,
 %\sout{ the red one the unstable regime with negative diffusion constant},
 %grey: %\sout{ the remaining grey one the}
 %strong coupling regime with the dimensionless coupling $g\to\infty$ in the hydrodynamic limit.}}
 }
 \label{cholesteric}
\end{figure}

%\sout{That is,}
  Thus the active
chiral flock %\sout{can be thought of as}
 resembles a cholesteric with the ``pitch
axis" being time; see Fig. \ref{cholesteric}. %\sout{This analogy is illustrated in Fig. \ref{cholesteric}.}}

\begin{figure}[htb]
\includegraphics[width=\columnwidth]{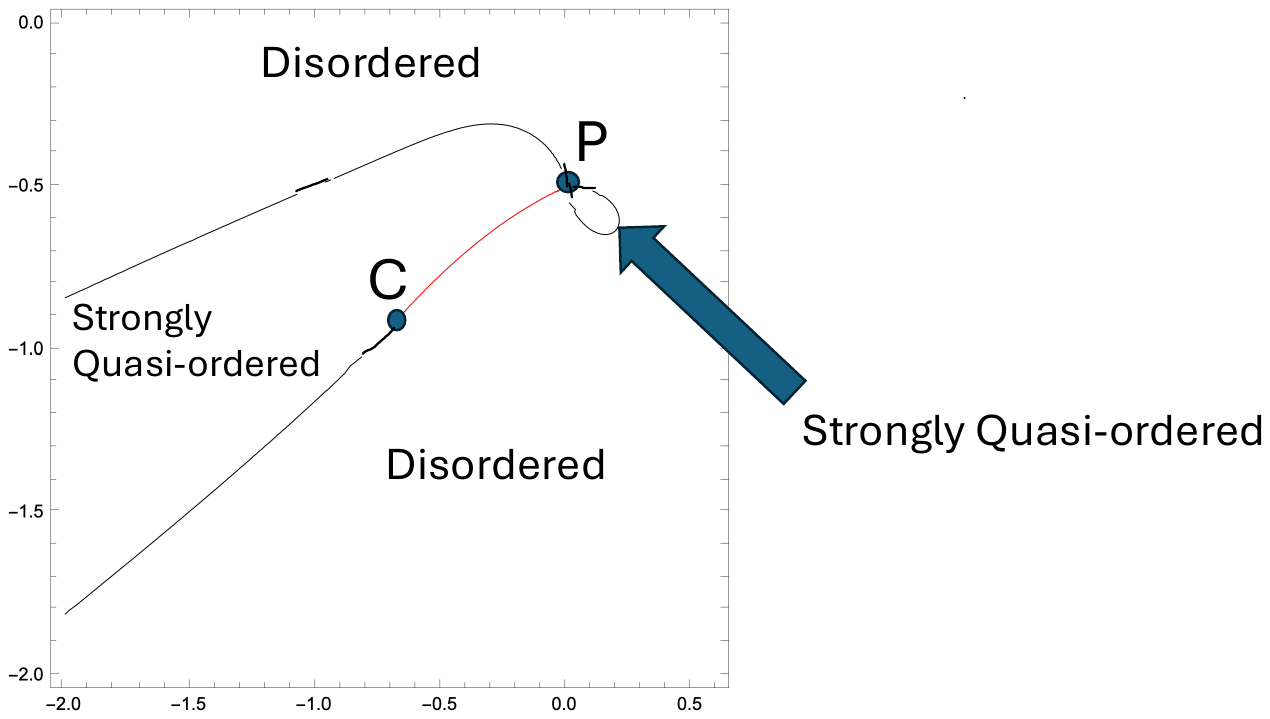}
 \caption{ Phase diagram of  chiral flocks in the $\mu_1$-$\mu_2$ plane for  modest values of $\mu_1$ and $\mu_2$.  Our strongly quasi-ordered phase occurs in two lobes which touch at the point $P$, where $\mu_1=-1/2, \mu_2=0$. The large lobe actually extends out to infinity in the negative $\mu_1$ and $\mu_2$ quadrant, in which its boundary is asymptotically bounded by the straight lines ${\mu_1\over\mu_2}=2$ and ${\mu_1\over\mu_2}=2.134$. The boundaries of these two regions are the loci of vortex unbinding phase transitions of the novel type described below, except for the portion of the boundary between the point $P$ and the point $C$ which lies at $\mu_1=-.92 \sep \mu_2=-.6872$. Along that section, the chiral flock disorders via a completely different type of continuous phase transition which we call the "instability" transition. See the text for a more detailed description.}\label{pd-small}
\end{figure}

\begin{figure}[htb]
\includegraphics[width=0.8\columnwidth,
height=10cm
]{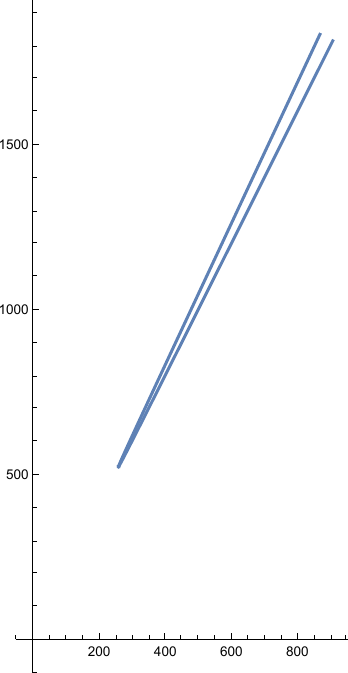}
 \caption{ Phase diagram of  chiral flocks in the $\mu_1$-$\mu_2$ plane for  large positive values of $\mu_1$ and $\mu_2$.  The "better ordered" phase occurs within the narrow sliver inside the curve, which also extends out to infinity in the negative $\mu_1$ and $\mu_2$ quadrant, in which its boundary is asymptotically bounded by the straight lines ${\mu_1\over\mu_2}=2$ and ${\mu_1\over\mu_2}=2.134$. The transition along this locus is a vortex unbinding transition, in the same universality class as that along the boundaries in figure \rf{pd-small}. Again, see the text for a more detailed description. }\label{pd-large}
\end{figure}

 \begin{figure}[htb]
 \includegraphics[width=0.9\columnwidth]{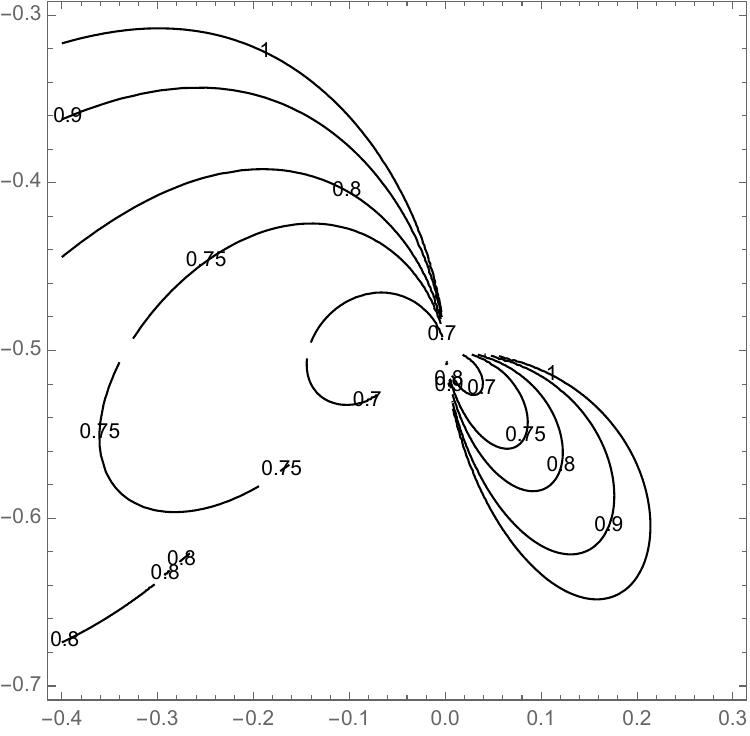}
 \caption{{ Contours of constant $\eta$ in the $\mu_1-\mu_2$ plane. These form two lobes of nested contour loops, with contours of smaller $\eta$ always lying inside those of larger $\eta$.   The size of the loops goes to zero as $\eta\to\left({2\over3}\right)^+$.    The contours with $\eta=1$ also form the phase boundaries of the two inner lobes of our "strongly quasi-ordered phase". }}\label{phase-lobes}
\end{figure}

In the presence of noise,  $\phi$ is no longer a constant. However, we find that there is a substantial range of parameters, which we will hereafter refer to as the ``strongly quasi-ordered phase'', over which $\phi$ is a slowly varying function of $\bf r$ and $t$. The fluctuations of $\phi(\br,t)$  diverge,  and,  as a result, the order parameter $|\langle \bv(\br,t)\rangle|$  decays as the spatial extent $L$ of the flock goes to infinity according to the law
\begin{eqnarray}
 &&\langle\phi^2({\bf r},t)\rangle = E \left[\ln \left({L\over a_0}\right)\right]^\eta \,,
 \label{phifluc}\\
 |\langle \bv(\br,t)\rangle|%&=&v_0 \exp\left[-{1\over2} \langle\phi^2({\bf r},t)\rangle\right]
%\nn\\
&=&v_0 \exp\left[-{E\over2}  \left(\ln \left(L\over a_0\right)\right)^\eta\right] \,,
\label{OPcor}
\end{eqnarray}
%{\cblue using $|\langle \bv(\br,t)\rangle|= v_0 \exp\left[ \langle\phi^2({\bf r},t)\rangle/2\right]$,}
 %\sout{with}
where the coefficient $E$, the microscopic length $a_0$,  and, most significantly, {\it the exponent} $\eta$ are all  {\it non-universal}  parameters of the flock. Though non-universal, $\eta$ is bounded below: $\eta > {2\over3}$.
%\beq
%\eta > {2\over3}\,.
%\label{etabound}
%\eeq
The fact that $ \langle\phi^2({\bf r},t)\rangle$ {\it does} diverge as $L\to\infty$ for all $\eta$'s allowed by \rf{etabound} implies  that 2D chiral flocks do {\it not} have long-range order, but can be infinitely { {\em more ordered }} (for $\eta<1)$  than  2D equilibrium XY models.

The sub-logarithmic decay of order implied by (\ref{OPcor}) for $\eta<1$ is so slow that it is not surprising that Ref.~\cite{Chate_chiral} found an ordered state like \rf{SSI}. Their system was, presumably, not big enough for them to observe the asymptotic decay of order implied by (\ref{OPcor}), since "big enough" could be very big indeed (indeed, it could be literally astronomical) if $E$ in (\ref{phifluc}) is small.

We derived the  result \rf{phifluc} in the so-called ``spin-wave" approximation~\cite{KT}, in which we neglect the possibility of topological defects - specifically, vortices~\cite{KT} - in the velocity field. %Including vortices is a formidable challenge in chiral flocks, which are fundamentally non-linear, so that the fields of vortices do not simply add. This precludes the possibility of treating them in anything like a "Kosterlitz-Thouless" theory.

We strongly suspect, and will present heuristic arguments to support this below, that the effects of vortices are negligible, at least for sufficiently small noise, in systems with $\eta<1$; that is, in chiral flocks that are {\it better} ordered than equilibrium XY models. The same arguments suggest that vortices are {\it always} important, even at arbitrarily small noise, if $\eta>1$. If this argument is correct, then the locus in parameter space on which $\eta=1$ is a phase transition boundary between { the strongly quasi-ordered phase} with phase fluctuations given by \rf{phifluc}, and a much less ordered ( specifically {\it short}-range ordered)  phase  riddled with a proliferation of unbound vortices. It also implies that the only $\eta$'s that will be observed in practice are those with $\eta$ bounded {\it both} below {\it and} above:
\beq
{2\over3}\le\eta\le1 \,.
\label{etabound}
\eeq

We find that number fluctuations are also unusual in chiral flocks. Specifically, if we consider an open region of such an area as to contain, on average, $\Nb$ active particles,  then% \sout{we find that}
the  standard deviation $\sigma(N)$ of the number of particles $N$ in this area scales with $\Nb$ as
\beq
\sigma(N)\propto \sqrt{\Nb/(\ln \Nb)^{1-\eta}} \,.
\label{minimum}
\eeq
From this, we see that, if $\eta<1$, as we suspect will always be the case, these fluctuations are always infinitely {\it less}, in the limit of mean number $\Nb\to\infty$, than the usual "law of large numbers" relation $\sigma(N)\propto \sqrt{\Nb}$. We therefore follow~\cite{diffusXY,diffusXY1}, who found the same phenomenon in diffusing active XY models, and refer to these by the appealing phrase  "miniscule number fluctuations" (hereafter "MNF"),  or ``hyperuniform''~\cite{hyperuniform}.% {\cred Alternatively, one could follow recent practice\cite{hyperuniform} and refer to these flocks as "hyperuniform".}

If our conjecture that regimes with $\eta>1$ are unstable against vortex proliferation is incorrect, then regions in which $\eta>1$ will have "Giant number fluctuations"   with $\sigma(N)$ being infinitely larger, in the limit of $\Nb\to \infty$, than the usual "law of large numbers" relation $\sigma(N)\propto \sqrt{\Nb}$, as can be seen directly from \rf{minimum}.

{ Even if we are correct that vortices always unbind for $\eta>1$,  in systems with small noise, there could be a wide range of length scales over which  giant number fluctuations could be seen.  { In that range of scales, ``weakly  quasi-ordered states'' with $\eta>1$ should be visible. }

Our theory also predicts the phase diagram shown in  Fig.~\rf{pd-small} and Fig.~\rf{pd-large}}.
The axes in that figure are two parameters  which we call $\mu_1$ and $\mu_2$ of the hydrodynamic theory that we will present
 below. Since we expect the parameters of our hydrodynamic theory to be smooth continuous  single-valued functions  of the control parameters in experiments or simulations on flocks, the {\it topology} of the phase diagram of a chiral active flock in the space of {\it any} two control parameters should have the topology of { Fig.~\rf{pd-small} and Fig.~\rf{pd-large}}.  Furthermore, if we tune just {\it one} experimental control parameter (call it $p$), since we expect $\mu_1$ and $\mu_2$ to both be functions of $p$, tuning $p$ will move the system along some locus in the $(\mu_1, \mu_2)$ plane.

There are three distinct types of phase transitions in our model:

\noindent 1) An ``instability'' transition occurs when the system crosses the  segment PC in figure \rf{pd-small}. As the transition is approached from the disordered side, by tuning an experimental control parameter which we will call $p$ (this is the analog of tuning temperature in an equilibrium system), orientational correlations decay  beyond  a correlation length $\xi$.  This correlation length  grows extraordinarily rapidly as the control parameter $p$ approaches its critical value $p_c$:
\begin{equation}
 \xi=a_1\exp \left[\ee^{\left(C/\sqrt{|s|}\right)}\right] \sep s\equiv |p-p_c|/p_c
 \label{instab corr len}
\end{equation}
where $a_1$  is a non-universal length and $C$  a non-universal dimensionless  constant.  %{\cred\sout{SM: Here, $p$ is a function of the two dimensionless parameters $\mu_1,\mu_2$; see below for precise definitions.}}

\noindent 2) { A "diffusion constant  driven" instability transition, which
 occurs when one tunes a third parameter: the ratio  $\gamma$ of the  diffusion
constant for  the density of active particles to the diffusion constant for the phase $\phi$. These diffusion constants are defined more
precisely below.  }

One can think of $\gamma$ as a third axis coming out of the plane of figure \rf{pd-small} or \rf{pd-large}. Defining the distance to the transition by analogy with that used for transitions of type (1):
%\beq
$s\equiv |\gamma-\gamma_c|/\gamma_c \,,$
%\label{sdef}
%\eeq
we find
that
 the associated correlation length also diverges strongly as $s\to0$ { from the disordered side}; specifically as
\begin{equation}
 \xi_c=a_2\exp\left({C_2\over s^\nu}\right)     \sep
 { s\equiv |\gamma-\gamma_c|/\gamma_c }
 \label{corr-2}
\end{equation}
where  $a_2$  is a constant, non-universal length and $C_2$  another non-universal dimensionless  constant. The exponent $\nu$  is also non-universal, and can vary from 0 to $\infty$.

\noindent 3) A vortex unbinding transition as $\eta$ passes
through $\eta=1$. Our analysis of this transition is more speculative and heuristic than the above two.  Making what we believe are reasonable assumptions, outlined below and discussed in details in the associated long paper (ALP)~\cite{alp},  we find that the correlation length in this case diverges extremely strongly as the transition is approached  from the disordered side: %. Specifically, we find
\begin{equation}
 \xi=a_v\exp\left[\ee^{\left(C_v/s\right)}\right]\sep
  \label{corr-3}
\end{equation}
where, as before, $s$ is the difference between whatever experimental parameter is being tuned to move through the phase diagram, an its value when the transition line $\eta=1$ is crossed. In addition,   $a_v$ (here ``$v$'' stands for ``vortex'') is a constant, non-universal length and $C_v$  another non-universal dimensionless  constant.

In addition to being the meeting point of three different phase transition lines, the very busy   point $P$ is also a sort of "accumulation point" for different values of $\eta$. More precisely, in both lobes, contours of constant $\eta$  for all values of $\eta$ in the allowed range $2/3<\eta<1$ meet at the point $P$. In fact, in the lower lobe, these contours are a set of nested closed loops, starting and ending at the point $P$, with the contours for smaller values of $\eta$ nested inside those for larger values. The size of these loops goes to zero linearly with the difference $\delta\equiv\eta-2/3$; see Fig.~\ref{phase-lobes}.

In the upper lobe, the structure is very similar; the  sole difference is that the contours do not completely close.  Instead, for $2/3<\eta<.999617$, they start at P, and end on the locus PC (not shown in figure Fig.~\ref{phase-lobes}), while
for $\eta$'s  in the narrow range  $.999617<\eta<1$  they extend out to reach $\mu_{1,2}\to-\infty$. For these values of $\eta$,  a second, disconnected branch comes in from infinity and ends on the locus PC.

In the disconnected outer sliver of strongly quasi-ordered phase shown in figure \rf{pd-large}, the contours of constant $\eta$ are a set of near hyperbolae, nested inside the boundary of that region, with the contours for smaller $\eta$'s again inside those of larger $\eta$.

We  derive the hydrodynamic equations for dry, immortal chiral flocks at large spatial and temporal scales. In an isotropic, homogeneous frictional medium ensuring no momentum conservation, the key slow variables are the local phase $\phi(\br,t)$ and local number density $c(\br,t)$ of the flockers of mean concentration $c_0$.
 Symmetry  specifically, rotation invariance and conservation law considerations imply that the hydrodynamic equations for $\phi(\br,t)$ and number fluctuations $\delta c(\br,t)\equiv c(\br,t)-c_0$ take the form
 \begin{eqnarray}
 %\begin{align}
  \frac{\partial\phi}{\partial t}&=&D_\phi\nabla^2\phi + \Omega_1\delta c+\frac{\lambda}{2} ({\boldsymbol\nabla}\phi)^2+ \Omega_2 (\delta c)^2 + f_\phi  \,,%\nn\\
  \label{phi-full-eq}\\
 \frac{\partial\delta c}{\partial t}&=&\Omega_3 \nabla^2\phi + D_c\nabla^2\delta c + \lambda_1 {\boldsymbol\nabla}\cdot(\delta c{\boldsymbol\nabla}\phi) \nonumber \\&+&\lambda_c \eps_{ij}(\partial_i\phi)(\partial_j \delta c)+ f_c  \, ,\label{c-full-eq}
 %\end{align}
\end{eqnarray}
 where, as usual in the spirit of hydrodynamic approaches, we have retained only the symmetry permitted terms  to lowest order in fields and gradients. The parameters $D_\phi, D_c$ are the  aforementioned phase and density diffusivities; $\lambda,\Omega_2,\lambda_1,\lambda_c$ are the nonlinear coupling constants in the theory, which may be individually positive or negative;
 $\Omega_1$ {  and $\Omega_2$ reflect the fact that the spontaneous rotation rate of the flockers can depend on the local density. The parameter $\Omega_3$ reflects the fact that particle currents can be driven by spatial variations of the phase $\phi$. The product $\Omega_1\Omega_3$ proves to be the square of the "speed of sound" $V_0$ of a pair of propagating coupled phase-concentration waves, which together are the hydrodynamic modes of chiral flocks.  L}inear stability requires  $\Omega_1\Omega_3>0$, which is the region on which we focus our attention here.

 In \rf{c-full-eq},   
 $\bm{\epsilon} $ is the anti-symmetric rank-2 tensor: $\epsilon_{11}=0=\epsilon_{22},\epsilon_{12}=1=-\epsilon_{21}$.
%\begin{equation}
%{\boldsymbol\epsilon} =
%\left(\begin{array}{rr}
%	0 & 1
%	\\
%	-1 & 0
%	\end{array}
%\right)\ .
%\label{epsdef}
%\end{equation}
The $\lambda_1-$ and $\lambda_c-$ terms in (\ref{c-full-eq}), respectively,  { contribute} to the particle current in response to a local phase difference along and perpendicular{ i.e., chiral} to ${\boldsymbol\nabla}\phi$.

 The additive noises $f_\phi(\br,t)$ and $f_c(\br,t)$ are Gaussian distributed, zero mean noises. Considering that $\phi(\br,t)$ and $\delta c(\br,t)$ are, respectively, nonconserved and conserved variables, the variances of $f_\phi$ and $f_c$ are chosen as
 \begin{eqnarray}
\langle f_\phi(\br,t)f_\phi(\br', t')\rangle = 2\Delta_\phi \delta^2(\br-\br') \delta(t-t')\,,\\
\label{facor}
\langle f_{c}(\br,t)f_{c}(\br', t')\rangle = -2\Delta_c\nabla^2 \delta^2(\br-\br') \delta(t-t')\,,
\label{fcscor}
\end{eqnarray}
where the noise strengths $\Delta_\phi,\Delta_c$ are phenomenological parameters of  the hydrodynamic theory. See ALP~\cite{alp} for more details.% derivation of the hydrodynamic equations.

In the linearized approximation, the equal-time correlations of $\phi$ and $\delta c$ in Fourier space can be calculated straightforwardly. In the long wavelength limit they are given by (with $\Gamma=(D_\phi+D_c)/2, \,\Delta=(\Delta_\phi+\Delta_c)/2$)
\begin{eqnarray}
 %C^0_{\phi\phi}(\bq)\equiv
 &&\langle |\phi({\bq},t)|^2\rangle_0 = \frac{\Delta}{\Gamma q^2} \sep%\label{theta-k-corr}\\
 %C_{cc}^0(k)\equiv
 \langle |\delta c({\bq},t)|^2\rangle_0 = \frac{\Delta}{\Gamma }\sep
 \label{corr_lin}
\end{eqnarray}
where %we have defined the mean diffusivity and noise strength
%\beq
%\Gamma=(D_\phi+D_c)/2 \,,~~~\Delta=(\Delta_\phi+\Delta_c)/2
%\label{gammadef}
%\eeq
the subscript ``0'' identifies  these as  linear theory results.

To go beyond the linear theory, we perform the perturbative dynamical Renormalization Group (hereafter "RG") on    Eqs.~(\ref{phi-full-eq}) and (\ref{c-full-eq}) to one-loop  order.  It turns out that only $D_\phi,\,D_c,\,\Delta_\phi$, and $\Delta_c$ receive relevant fluctuation-corrections at one-loop order due to the nonlinear terms. These corrections are $\log$-divergent in 2D. However, there are {\em no} diverging fluctuation corrections to  $\Omega_1,\Omega_3$ or  $\lambda,\,\Omega_2,\,\lambda_1,\,\lambda_c$ at one-loop order.  Furthermore, none of the relevant fluctuation corrections of $D_\phi,D_c,\Delta_\phi,\Delta_c$ involves $\lambda_c$.

We give  the details of this RG calculation in the ALP \cite{alp},  where we also show that our one-loop results become asymptotically exact in the long-distance, long-time limit.

The  one-loop differential RG recursion relations for $D_\phi,D_c,\Delta_\phi,\Delta_c$ are
 \begin{eqnarray}
  \frac{\dd D_\phi}{\dd\ell}&=& \frac{\lambda^2}{64\pi\Gamma^3}\{(D_c\Delta_\phi -D_\phi\Delta_c)\nonumber \\&+& [(2D_\phi+D_c)\Delta_c +D_\phi\Delta_\phi]\mu_2\} (1+2\mu_1)  \,,\label{flow-dh}\\
  \frac{\dd D_c}{\dd\ell}&=& \frac{\lambda^2}{32\pi\Gamma^3}(D_\phi+D_c) (2\mu_1\Delta_c+\mu_2\Delta_\phi)\mu_2  \,,
  \label{flow-dc}\\
 \frac{\dd\Delta_\phi}{\dd\ell}&=& \frac{\lambda^2(\Delta_\phi+\Delta_c)^2}{64\pi\Gamma^3}\left(1+4\mu_1+4\mu_1^2\right)  \,,\label{flow-delh}\\
  \frac{\dd\Delta_c}{\dd\ell}&=&\frac{\lambda^2\mu_2^2(\Delta_\phi+\Delta_c)^2}{32\pi\Gamma^3} \,,\label{flow-delc}
 \end{eqnarray}
where  we have defined the  two dimensionless parameters
\begin{equation}
\mu_1=\frac{\Omega_2}{\lambda},\;\, \mu_2=\frac{\lambda_1}{\lambda}.\label{mu-defn}
 \end{equation}
 Since $\lambda,\,\Omega_2$, and $\lambda_1$ do not renormalize at one-loop order,  the ratios $\mu_1,\mu_2$ do not either.

 We will show later that this failure to renormalize at one-loop order %implies that, for some range of bare (i.e., before renormalization) ratios $\mu_{1,2}$, those ratios flow, as RG time goes to infinity, to non-universal, finite constants. That is,
 rather than having a universal fixed point in the $\mu_{1,2}$ plane (which is the sort of behavior usually found under the RG), %that plane (or, more precisely, a portion
%of that plane)
 means that a portion of the $\mu_1$-$\mu_2$ plane, in which the strongly quasi-ordered phase exists, is a {\it fixed plane} under the RG. This is analogous {\it in the sense of being a fixed plane, rather than a single fixed point},  to the well-known~\cite{KT} "fixed line" that controls the Kosterlitz-Thouless ordered phase of the equilibrium  two-dimensional XY model. As mentioned above, however, the properties of the system in the strongly quasi-ordered  phase controlled by this fixed plane are quite different from those of the Kosterlitz-Thouless ordered phase.

  We further define the additional dimensionless ratios:
 %\begin{equation}
  $\gamma=\frac{D_c}{D_\phi},\;\alpha\equiv\frac{\Delta_c}{\Delta_\phi} \, $.% \label{dim-1}  \,.
 %\end{equation}
  $\,$ These {\it do} renormalize. Their recursion relations can be derived from the full recursion relations (\ref{flow-dh})-(\ref{flow-delc}), combined with the fact that $\mu_1$ and $\mu_2$ do not renormalize. % Lastly, we define $\Gamma=(D_\phi+D_c)/2,\Delta = \Delta_\phi +\Delta_c$.
   They imply that $\gamma(\ell)$ and $\alpha(\ell)$ flow, in the region  in  Fig.~\rf{pd-small} and Fig.~\rf{pd-large} of the $\mu_1$-$\mu_2$ plane occupied by  the strongly quasi-ordered  phase, to fixed point values $\alpha^*(\mu_1, \mu_2)$ and $\gamma^*(\mu_1, \mu_2)$ which depend {\it only} on the values of $\mu_1$ and $\mu_2$. (The rather complicated analytic expressions for $\alpha^*(\mu_1, \mu_2)$ and $\gamma^*(\mu_1, \mu_2)$ in terms of $\mu_1$ and $\mu_2$ are given in the ALP~\cite{alp}.

The  one-loop perturbation theory is  accurate when the effective dimensionless coupling constant  $g\equiv \frac{\lambda^2\Delta_\phi}{64\pi\Gamma^3}\, $  is small.

  Once $\alpha$ and $\gamma$ have flown to their fixed point values  $\alpha^*(\mu_1, \mu_2)$ and $\gamma^*(\mu_1, \mu_2)$,  the recursion relations \rf{flow-dh}-(\ref{flow-delc})  imply that $g$ obeys
 a closed recursion relation
  \begin{eqnarray}
 \frac{\dd g}{\dd\ell}&=&g^2 (\beta_{_\Delta}-3\beta_{_\Gamma})=-g^2 \beta_g(\mu_1,\mu_2) \,,
 \label{g-flow-A}
\end{eqnarray}
where $\beta_{_\Delta}(\mu_1,\mu_2)$ and $\beta_{_\Gamma}(\mu_1,\mu_2)$ are  defined by the equally closed recursion relations
 for
 the mean noise  $\Delta$ and mean damping $\Gamma$:
  \begin{eqnarray}
 \frac{\dd \Delta}{\dd\ell}=g\beta_\Delta(\mu_1,\mu_2)\Delta \sep
   \frac{\dd \Gamma}{\dd\ell}&=&g\beta_\Gamma(\mu_1,\mu_2)\Gamma \,.
  \label{Delta-Gamma-flow-A}
\end{eqnarray}
Equations \rf{g-flow-A} and \rf{Delta-Gamma-flow-A} follow from (\ref{flow-delh}) and (\ref{flow-delc}),  after using  the fixed point values $\gamma^*$ and $\alpha^*$ defined above. Detailed (and complicated!) analytic expressions for $\beta_\Delta$,  $\beta_{_\Gamma}$ and $\beta_g$ are given in the ALP~\cite{alp}.

We focus on the case $\beta_g(\mu_1,\mu_2)>0$ (i.e., $\beta_{_\Gamma}>\beta_\Delta/3>0$), because in this regime, which is a well defined  and non-vanishing  portion of the $\mu_1$-$\mu_2$ plane,   $g(\ell\to\infty)\to0$.  This implies that our perturbation theory,  which is valid for small $g$, becomes asymptotically exact in this regime.  In this regime,  the solution of \rf{g-flow-A} for large RG "time" $\ell$ is
\beq
g(\ell)\approx 1/(\beta_g\ell)
\label{gsol}
\eeq
 Using this  in \rf{Delta-Gamma-flow-A}, we find, again for large $\ell$,
\begin{eqnarray}
&&\Delta(\ell)=\Delta_0 \ell^{\eta_{_\Delta}},\,\Gamma(\ell)=\Gamma_0\ell^{\eta_{_\Gamma}},\label{Delta_ell}\\
\end{eqnarray}
where
\begin{eqnarray}
 &&\eta_{_\Delta}\equiv { \beta_{_\Delta}(\mu_1,\mu_2)\over3 \beta_{_\Gamma}(\mu_1,\mu_2)- \beta_{_\Delta}(\mu_1,\mu_2)}>0,
\label{etaD}\\
&&\eta_{_\Gamma}\equiv  { \beta_{_\Gamma}(\mu_1,\mu_2)\over3 \beta_{_\Gamma}(\mu_1,\mu_2)- \beta_{_\Delta}(\mu_1,\mu_2)}>0.
\label{etaG}
\end{eqnarray}
Here, $\Delta_0$ and $\Gamma_0$ are non-universal constants.   As usual in the RG, we can define an "effective" values  $\Delta(\bq)$ and $
\Gamma(\bq)$ of the noise and damping at wavevector $\bq$.  These have the property that  the true
two-point equal-time Fourier transformed correlation functions $\langle |\phi (\bq,t)|^2\rangle, \langle |\delta c (\bq,t)|^2\rangle$ at
wavevector $\bq$ of the full non-linear system are given by the expressions (\ref{corr_lin}) obtained from the linear theory,
except that everywhere the noise strength $\Delta$ and damping $\Gamma$ appear in those linear expressions, they must be
replaced by $\Delta(\ell^*(\bq))$ and $\Gamma(\ell^*(\bq))$ respectively. Here
%\beq
$\ell^*(\bq) = \ln (\Lambda/q)$ % \,,
%\label{ell*}
%\eeq
with $\Lambda$   a phenomenological ultraviolet cutoff  (i.e., an upper limit on $q=|\bq|$) of order the inverse of some microscopic length (e.g., the inter-flocker distance).  Using this value of $\ell^*$  in (\ref{Delta_ell}) and the resultant $\Delta(q)$ and $\Gamma(q)$ in (\ref{corr_lin}), we { obtain} the renormalized correlations  (which are the ones observed in experiments) are given by
\begin{eqnarray}
 &&\langle |\phi({\bf q},t)|^2\rangle = \frac{\Delta(q)}{\Gamma(q) q^2}=\frac{\Delta_0}{\Gamma_0}\left[\ln \left(\frac{\Lambda}{q}\right)\right]^{\eta_{_\Delta}-\eta_{_\Gamma}}\frac{1}{q^2}\,,
 \nn\\
 \label{renorm-cphi}\\
 &&\langle |\delta c({\bf q},t)|^2\rangle = \frac{\Delta(q)}{\Gamma(q)}=\frac{\Delta_0}{\Gamma_0}\left[\ln \left(\frac{\Lambda}{q}\right)\right]^{\eta_{_\Delta}-\eta_{_\Gamma}}\,.
 \label{renorm-cc}
\end{eqnarray}

The real space phase fluctuations  in a system of linear spatial extent $L$
can be calculated by Fourier transforming (\ref{renorm-cphi}) back to real space:
\begin{eqnarray}
 \langle \phi^2(\br,t)\rangle%&=&\int_{q=1/L}^{q=\Lambda} \frac{\dd^2q}{(2\pi)^2}\frac{\Delta_0}{\Gamma_0}\left[\ln \left(\frac{\Lambda}{q}\right)\right]^{\eta_{_\Delta}-\eta_{_\Gamma}}\frac{1}{q^2}
 %\nonumber \\
 %&=&\frac{\Delta_0}{\Gamma_0}\int_{1/L}^\Lambda \frac{\dd q}{(2\pi)}\left[\ln \left(\frac{\Lambda}{q}\right)\right]^{\eta_{_\Delta}-\eta_{_\Gamma}}\frac{1}{q}
 %\nonumber \\
 &=& E\left[\ln \left(\frac{L}{a_{_0}}\right)\right]^\eta  \,,\label{renorm-cphi-real}
 %&&\eta\equiv 1+\eta_{_\Delta}-\eta_{_\Gamma} \,,
\label{etadef}
\end{eqnarray}
where we have defined
\beq
\eta\equiv 1+\eta_{_\Delta}-\eta_{_\Gamma} \,.
\label{etaval}
\eeq
 In \rf{etadef}, $E\equiv \Delta_0/2\pi\eta\Gamma_0$,    and the $q$-integration is from $q=1/L$ to $1/a_{_0}$. Since both $ \eta_{_\Gamma}$ and $\eta_{_\Delta}$ vary continuously with $\mu_1$ and $\mu_2$, whose values vary from system to system,  the exponent $\eta$ is not universal.  Therefore, the real-space phase fluctuations (\ref{renorm-cphi-real}) imply {\em continuously varying logarithmic}  growth of fluctuations. This is quite an unusual feature of this problem; usually~\cite{chaikin}, scaling laws are universal for all systems of a given symmetry.

The variability of $\eta$ has limits, however.   In the region of the $\mu_1$-$\mu_2$ plane we are considering (which is the only regime in which equation \rf{renorm-cphi-real}
applies)  $\eta$ is bounded below. To see this,  we use  our expressions \rf{etaD} and \rf{etaG} for $\ed$ and $\etg$ in {\rf{etaval}},  to obtain
\beqn
\eta&=&{\beta_g+\beta_{_\Delta}-\beta_{_\Gamma}\over\beta_g}%={3\beta_{_\Gamma}-\beta_{_\Delta}+\beta_{_\Delta}-\beta_{_\Gamma}\over3\beta_{_\Gamma}-\beta_{_\Delta}}\nn\\
={2\beta_{_\Gamma}\over3\beta_{_\Gamma}-\beta_{_\Delta}}%={2\over3-{\beta_{_\Gamma}\over\beta_{_\Delta}}}
>{2\over3} \,,
%\label{etabound}
\eeqn
where the last inequality follows from the fact that, in the regime of $\mu_{1,2}$ we are discussing, ${\beta_{_\Gamma}\over\beta_{_\Delta}}$ is positive, since $\beta_{_\Gamma}$ and $\beta_{_\Delta}$ both are.

There is also an {\it upper} bound on $\eta$; specifically, $\eta\le1$. This bound holds because systems with $
\eta>1$ are unstable to the unbinding of vortices - that is, topological defects in the phase field $\phi$, like those that disorder the
equilibrium 2D XY model at sufficiently high temperatures\cite{KT}.

This bound can be understood as arising from the fact that for $\eta>1$, the "effective reduced temperature" of a chiral flock renormalizes to infinity, thereby passing the Kosterlitz-Thouless threshold for vortex unbinding.

To make this idea more precise, we note that, if we compare our result
\rf{etadef} for the $\phi$ fluctuations in real space with the analogous result\cite{KT} 
 \beq
 \langle |\phi({\bf r},t)|^2\rangle =  \left({k_BT\over 2\pi K}\right)\ln\left({L\over a_{_0}}\right)\,.
\label{ktfluc}
\eeq
$\,$

\noindent for the equilibrium XY model (whose Goldstone mode variable is also a phase $\phi$),
we see that the two are equivalent if we define an effective, length scale dependent ``reduced temperature'' $T_{\rm eff \, reduced}$ given by
%\bew
 \beq
T_{\rm eff \, reduced}(L)\equiv\left({k_BT\over 2\pi K}\right)_{\rm eff}(L) = E\left[\ln \left(\frac{L}{a_{_0}}\right)\right]^{\eta-1}\,.
 \label{Teff1}
 \eeq
% \ew
 %{\cred\sout{where in the last equality, we have used the result \rf{etaval} for $\eta$.}}
 Thus we see that the effective reduced temperature diverges as $L\to\infty$ if $\eta>1$ and vanishes as $L\to\infty$ if $\eta<1$.
 The former result implies that vortices will always unbind when $\eta>1$, since they do\cite{KT} for sufficiently large  $T_{\rm eff \, reduced}$. Thus, we can only find a  strongly quasi-ordered phase , whose existence depends on the absence of vortices, for $\eta\le1$. Indeed, the locus $\eta=1$ in the $\mu_1$-$\mu_2$ plane is the locus of a vortex unbinding transition at which the order of the strongly quasi-ordered  phase disappears. More on this later.

%\sout{At first glance, it would appear that these fluctuations could
%be {\it bigger} than those of  the 2D equilibrium XY model (when $\eta>1$) or  {\it smaller} (when $\eta<1$). However, we argue in
%the ALP that, in fact, there is also an {\it upper} bound on $\eta$; namely, $\eta\le1$.}

 Hence  for $\eta\le1$ the fluctuations in this system are always smaller than, or comparable to, those of the equilibrium XY model. Note, however, that, since our {\it lower} bound on $\eta$ is greater than $0$, the fluctuations always diverge in the limit of system size $L\to\infty$. Hence, the system never has true long-ranged orientational order, even though it is generically  {\it better} ordered than the  well-known QLRO of the 2D equilibrium XY model.

Next we calculate the number fluctuation correlations in the real space and the variance $\langle (\delta N)^2\rangle$ of the number fluctuations in an open finite square area of size $\tilde L\times \tilde L$.  Inverse Fourier transforming (\ref{renorm-cc}), we find
\begin{equation}
 C_{cc}^R(r) \equiv \langle\delta c(\bR, t)\delta c(\bR+\br, t)\rangle
 \sim \frac{ D_0}{2{\Gamma_0}}\frac{[\ln (\Lambda r)]^{\eta-1}}{2\pi r^2} \label{ren-c-corr}
\end{equation}
for large $r$. %{\cmag
%\sout{From (\ref{ren-c-corr}) we find that
%\begin{equation}
% C_{cc}^R(r)|_{\eta<1}\ll C_{cc}^R(r)|_{\eta >1}
%\end{equation}
%for large $r$, which means that the number density fluctuations are strongly suppressed for $\eta<1$ (MNF or hyperuniformity) in compared to $\eta\geq 1$  in the limit of large $r$. In fact,   the number density fluctuations are progressively suppressed as $\eta$ decreases. }}
Then, as shown in ALP~\cite{alp},
\bew
\begin{eqnarray}
\langle (\delta N)^2\rangle &=& \int_0^{\tilde L}d^2r \int_0^{\tilde L} d^2r'
\,C_{cc}^R(\br-\br')
%\int d^2k \int d^2 k' e^{i{\bf k}\cdot {\bf r}}e^{i{\bf k'}\cdot {\bf r'}}
%\nonumber \\
%\times\langle \delta n ({\bf k},t)\delta n ({\bf k'},t)\rangle
%&&\sim \int_0^{\tilde L} d^2R' \int_0^{\tilde L} d^2 R \exp[i{\bf k}\cdot ({\bf r-r'})]\langle |\delta n({\bf k},t)|^2\rangle\nonumber \\
%&\sim&
\sim\langle N\rangle
\left(\ln \langle N\rangle\right)^{(\eta-1)}\,,\nn%\label{numb-fluc}
\end{eqnarray}
\ew
which leads to  our result (\ref{minimum}) for $\sigma(N)\equiv\sqrt{\langle (\delta N)^2\rangle}$.  This means that the number fluctuations are strongly suppressed  { for $\eta<1$} compared to those in equilibrium systems.    We call this ``miniscule number fluctuations'' (MNF),  or ``hyperuniformity''.

The recursion relations (\ref{flow-dh})-(\ref{flow-delc}) above can also be used to show that there are three distinct types of phase transitions { out of } the { strongly quasi-ordered } phase in chiral flocks.

 The first of these is the ``instability transition'', which occurs along the segment PC in figure \rf{pd-small}. This transition occurs when $\mu_1$ and/or  $\mu_2$ are tuned into a region in which   the stable fixed point $\gamma^*$ for $\gamma$  disappears. When this happens,  the system flows under renormalization into a regime of negative diffusion constants, in which the ordered state becomes dynamically unstable.  In the ALP~\cite{alp},  we explain in detail how we obtain this locus numerically, and how we analytically obtain  the associated correlation length (\ref{instab corr len}).

%From our analytic expressions for $a$, $b$, and $c$ in terms of $\mu_1$ and $\mu_2$ above, we can, with  help from our friend Little Stevie Wolfram, identify this regime
%in the $\mu_{1,2}$ plane. It is the regime inside the figure eight curve shown in figure \rf{pd}; that curve is simply the locus on which $b^2 - 4ac=0$. If an experimental system is "tuned" through this locus, it therefore undergoes the order-disorder transition of the type "1" described above

 The second, ``diffusion constant driven'', transition occurs if $\mu_1$ and $\mu_2$ are in the region in which a stable fixed point { $\gamma_s$} for $\gamma$ {\it does} exist, but
 the bare value of the parameter $\gamma$ is on the "wrong" side of   an {\it unstable}  value $\gamma_u$;  that is, on the side that flows away from {\it both} $\gamma_u$ and $\gamma_s$. { We show in the ALP that such a point {\it always} exists for any value of
$\mu_1$ and $\mu_2$ such that a stable fixed point for $\gamma$ {\it does} exist.

When this happens, }
 then the system also flows into a regime in which the ordered state becomes dynamically unstable. Tuning $\gamma$ from the ``right'' to the ``wrong'' side of $\gamma_u$ therefore leads to another type of order-disorder transition. This is the transition of the type ``2'' described above.

 The third  transition, which is the vortex unbinding transition, can, as we explained earlier,  be understood as being driven by the renormalization of the ``effective temperature'' to infinity, which  occurs when the exponent $\eta$ increases through $1$.  As also discussed above, this results   in the unbinding of vortices.  The recursion relations  \eqref{flow-dh}-\eqref{flow-delc}  and their analysis up to now has ignored the possibility of vortices. Including them systematically is an extremely difficult problem, since the system is non-linear. As a result,   the fields of vortices, in contrast to those of the equilibrium XY model, do not simply add. This makes the problem much more difficult.

   However, our  earlier  intuitive argument,  based on identifying $T_{\rm eff \, reduced} (L){ \propto L^{\eta-1}}$ as the {\em scale-dependent}  analog of the reduced temperature $k_BT/K$ in the 2D XY model  strongly suggests that when $\eta>1$; vortices {\it do} unbind. Consequently, only short range order is expected. In contrast for $\eta<1$, $T_{\rm eff \, reduced} \rightarrow 0$ for $L\rightarrow \infty$, suggesting a vortex-free ordered state, at least for low noises.  %This argument can be more formally stated by adding to our  recursion relations for the chiral flock without vortices,  the recursion relations for the equilibrium XY model {\it with} vortices; see ALP~\cite{alp}.}
%
%\sout{We have instead included them}

 To address this transition more quantitatively, in the ALP~\cite{alp}, we use  a ``kluge'': we   add to our  recursion relations for the chiral flock without vortices terms from  the recursion relations for the equilibrium XY model {\it with} vortices. More details are given in the ALP~\cite{alp};
the resultant recursion relations lead to the prediction  \eqref{corr-3} for the behavior of the correlation length $\xi_v$. Obviously, that result is somewhat speculative, in contrast to our results for the other two transitions, of which we are quite confident.

In summary, we have shown that chirality radically changes flocking in two dimensions. It reduces the long-ranged order present in achiral flocks to a type of order that is better than the quasi-long-ranged order found in the equilibrium 2DXY model, but still not long ranged. { Nonetheless, this order is {\em infinitely} stronger than the short-range order found in mortal or Malthusian chiral flock~\cite{JT-chiral-pre}.} This order is characterized, as illustrated in (\ref{phifluc}), by a non-universal exponent  $\eta$ giving the scaling of the decay of orientational order with the logarithm of the system size;  $\eta=1$ always in the equilibrium XY model, but here varies continuously between $2/3$ and $1$.  The number density fluctuations are also suppressed in these nearly ordered states, giving hyperuniformity,  being characterized by the {\em same} non-universal exponent $\eta$. We have also found three different types of phase transitions out of this nearly ordered state, all of which belong to heretofore undiscovered universality classes, whose universal properties we have worked out. { Number density fluctuations  help to stabilize order in immortal chiral flocks; { this} mechanism is absent in Malthusian chiral flocks, { which are therefore disordered~\cite{JT-chiral-pre}.  }}%{\cblue [LC: I agree with John's suggestion. Let's be vague.]}

A.B. and S.M. thank Alexander von Humboldt Stiftung (Germany)
for partial financial support through their research group linkage programme (2024). A.B. thanks ANRF (India) for partial financial
support through the ARG (MATRICS) programme (file no.:
ANRF/ARGM/2025/000461/TS). S.M. thanks ANRF (India) for partial financial
support through the ARG (MATRICS) programme (file no.: ANRF/ARGM/2025/000748/TS). JT thanks the Max-Planck Institut f\"ur Physik Komplexer Systeme, Dresden, and the Higgs Center at the University of Edinburgh, for their hospitality and support while much of this work was underway.

\end{document}